\documentclass[aps,pre,amsmath,amssymb,twocolumn]{revtex4-2}
\usepackage{graphicx}
\usepackage{dcolumn}
\usepackage{float} 
\usepackage{bm}
\usepackage[utf8]{inputenc}
\usepackage[T1]{fontenc}
\usepackage{physics}
\usepackage{amsmath}
\usepackage{placeins}
\usepackage{float}
\usepackage{color}
\usepackage[most]{tcolorbox}
\usepackage{tikz}
\usepackage{relsize}
\usepackage[normalem]{ulem}
\usepackage{parskip}
\usepackage{xr}

\begin{document}


\title{Emergent turbulence and coarsening arrest in active-spinner fluids}

\author{Biswajit Maji\textsuperscript{1,\#}}
\email{biswajitmaji@iisc.ac.in}

\author{Nadia Bihari Padhan\textsuperscript{2,\#}}
\email{nadia_bihari.padhan@tu-dresden.de}

\author{Rahul Pandit\textsuperscript{1}}
\email{rahul@iisc.ac.in}

\affiliation{
\textsuperscript{1}Centre for Condensed Matter Theory, Department of Physics, Indian Institute of Science, Bangalore 560012, India \\
\textsuperscript{2}Institute of Scientific Computing, TU Dresden, 01069 Dresden, Germany
}

\thanks{\textsuperscript{\#}These authors contributed equally to this work.}







\begin{abstract}
We uncover activity-driven crossover from phase separation to a new turbulent state in a two-dimensional system of counter-rotating spinners. We study the statistical properties of this active-rotor turbulence using the active-rotor Cahn-Hilliard-Navier-Stokes model, and show that the vorticity  $\omega \propto \phi$, the scalar field that distinguishes regions with different rotating states. We explain this intriguing proportionality theoretically, and we characterize power-law energy and concentration spectra, intermittency, and flow-topology statistics. We suggest biological implications of such turbulence.

\end{abstract}

\maketitle

\section{Introduction} Turbulence abounds in nature: it manifests itself from astrophysical to biophysical scales and continues to pose challenging problems for physicists, mathematicians, biologists, and engineers. Over the last decade or so, turbulence in active fluids, driven via internal active mechanisms and not by external energy input, has attracted a lot of attention [see, e.g., Refs.~\cite{wensink2012meso,bratanov2015new,thampi2016active,saintillan2018rheology,rana2020coarsening,bowick2022symmetry,sanjay2022transport,alert2022active,gibbon2023analytical,kiran2023irreversibility,mukherjee2023intermittency,padhan2024novel,padhan2024interface,rana2024defect,kiran2025onset}]. Such active fluids belong to the class of nonequilibrium  active-matter systems~\cite{ramaswamy2010mechanics,fodor2016far} that include the collective motion of self-propelled particles~\cite{vicsek2012collective}, liquid phases without attractive forces~\cite{tailleur2008statistical}, and dense bacterial suspensions and their non-living equivalents~\cite{aranson2013active,driscoll2019leveraging,zottl2016emergent,elgeti2015physics}, to name but a few. There is increasing interest in active-fluid systems with self-rotating particles~\cite{yeo2015collective,nguyen2014emergent,goto2015purely,kokot2017active}, which have been shown to display a new type of torque-induced phase separation~\cite{zhang2021active, dauchot:hal-03363365}. In particular, mixtures of counter-rotating particles exhibit phase separation in both \textit{wet}~\cite{fily2012cooperative} and \textit{dry}~\cite{riedel2005self} environments. Studies of self-rotating particles have important biological implications for they are found in a variety of biological systems, e.g., spinning organisms like sperm-cell clusters~\cite{riedel2005self}, the bacterium \textit{Thiovulum majus}~\cite{petroff2015fast}, and dancing \textit{Volvox} algae~\cite{drescher2009dancing}; synthetic experimental systems have also been developed~\cite{kokot2018manipulation,soni2019odd,han2020emergence}.

\begin{figure*}[htp]
  \includegraphics[width=1.0\textwidth ]{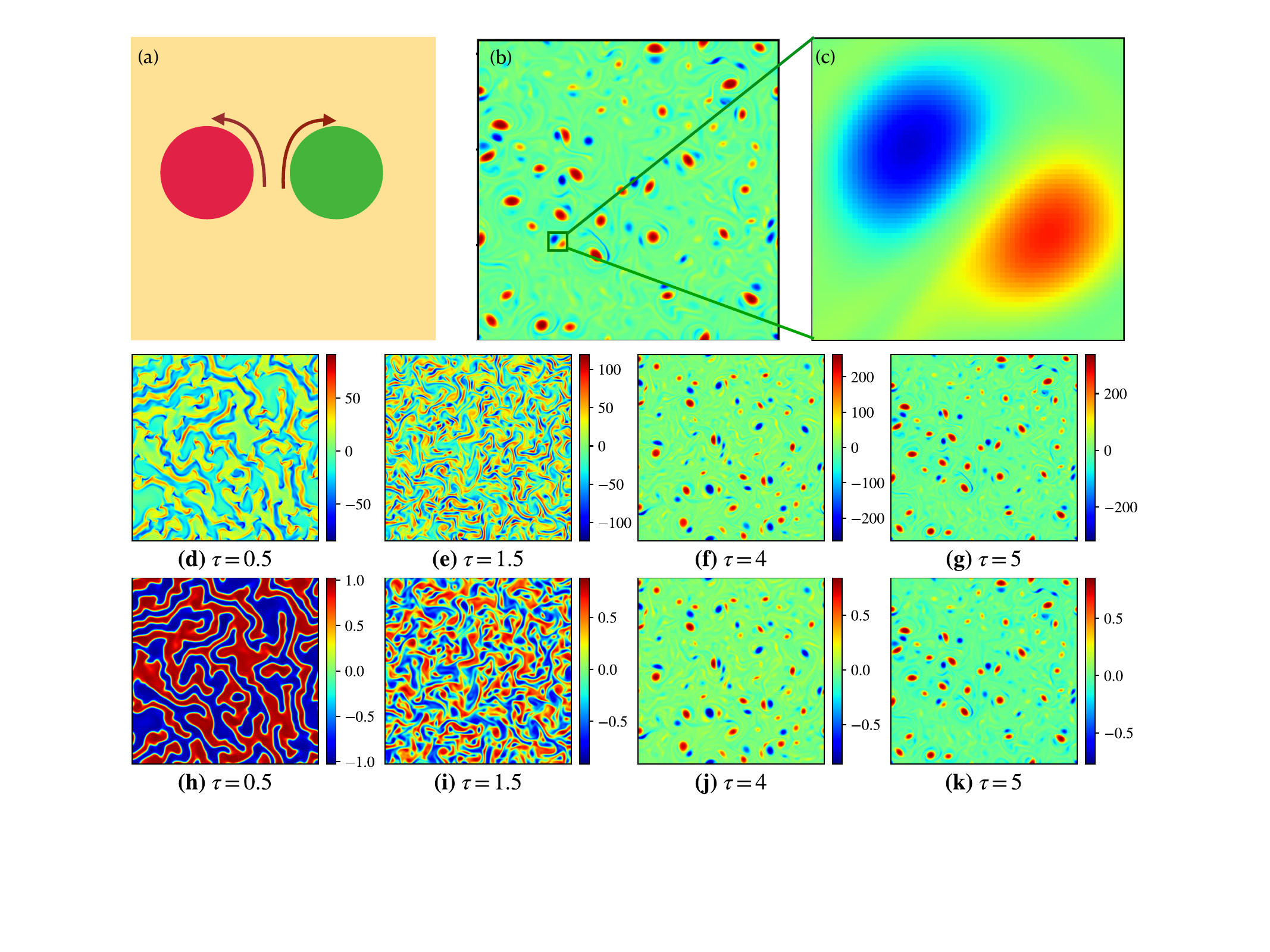}  
\caption{(a) A schematic diagram of counter-rotating rotors; (b) an illustrative pseudocolor plot of $\omega$ in the active-rotor-induced turbulent NESS with
vortex doublets [shown enlarged in (c)]; pseudocolor plots of $\omega$ for (d) $\tau=0.5$, (e) $\tau=1.5$, (f) $\tau=4$, and (g) $\tau=5$, with corresponding pseudocolor plots for $\phi$ in (h), (i), (j), and (k), respectively, at representative times, showing partial phase separation at low values of $\tau$ and turbulent NESSs, with $\omega \propto \phi$, at large values of $\tau$. As $\tau$ increases, the number of vortex doublets rises, but their sizes decrease.
For the full spatiotemporal evolution of these fields see Videos V1-V4 in the Appendix.}
\label{fig:pcolor}
\end{figure*}
\begin{figure*}[htp]
 \includegraphics[scale=0.70]{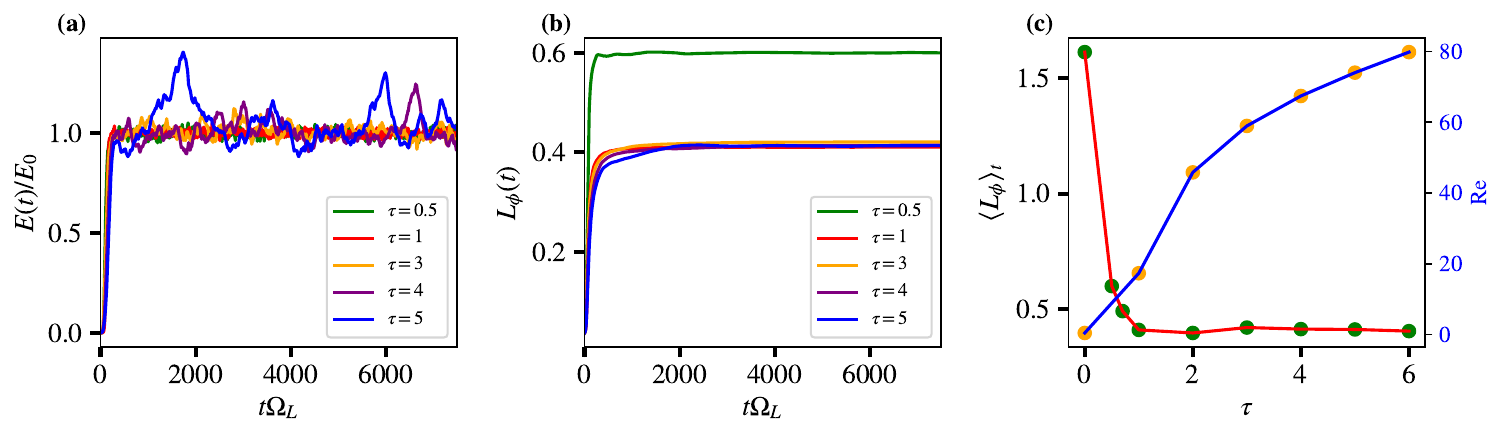}
\caption{Plots versus the scaled time $t\Omega_L$ of (a) the scaled energy $E(t)/E_0$, where $E_0=E(t=\Omega_L^{-1})$, and (b) the coarsening-arrest length scale $L_\phi(t)$, for $\tau = 0.5, 1, 1.5, 4$, and $5$. (c) Plots versus $\tau$ of the mean domain size $\langle L_\phi \rangle$ (red curve) and the integral-scale Reynolds number $Re$ (blue curve).}
 \label{fig:length_scale}
\end{figure*}

We consider a system of counter-rotating active spinners [Fig.~\ref{fig:pcolor} (a)] that exhibits coarsening~\cite{sabrina2015coarsening}, into phases with clockwise- and anticlockwise-rotating spinners. We demonstrate that, as we increase the activity, this system crosses over to a hitherto unanticipated  nonequilibrium statistically state (NESS) of \textit{emergent active-rotor turbulence}, which leads, in turn, to coarsening arrest. We examine the statistical properties of active-rotor turbulence, both theoretically and numerically, and show that it is markedly different from two-dimensional (2D) fluid turbulence~\cite{pandit2017overview,boffetta2012two} and active-fluid turbulence in a variety of systems [see, e.g., Refs.~\cite{wensink2012meso,bratanov2015new,sanjay2022transport,alert2022active,kiran2023irreversibility,mukherjee2023intermittency,padhan2024novel,padhan2024interface,kiran2025onset}]. We discuss the possible implications of our work for biological and synthetic systems with active spinners~\cite{grzybowski2000dynamic,drescher2009dancing,kokot2018manipulation,soni2019odd,han2020emergence}.

\section{Model and Methods}
The minimal hydrodynamical description of active-rotor system~\cite{sabrina2015coarsening} uses the following partial differential equations (PDEs) [henceforth, the active-rotor Cahn-Hilliard-Navier-Stokes (ARCHNS) model] in 2D  in terms of the vorticity $\bm{\omega}$ [for the velocity formulation of Eq.~\eqref{eq:omega} see the Appendix]:
\begin{eqnarray}
    \partial_t \phi + (\bm u \cdot \nabla) \phi &=& M \nabla^2 \left( \frac{\delta \mathcal F}{\delta \phi}\right)\,; \label{eq:phi}  \\
     \mathcal F[\phi, \nabla \phi] &=& \int_{\Omega} \left[\frac{3}{16} \frac{\sigma}{\epsilon}(\phi^2-1)^2 + \frac{3}{4} \sigma \epsilon [\nabla \phi]^2\right]\,;\label{eq:functional}\\
    \partial_t \omega + (\bm u \cdot \nabla) \omega &=& \nu \nabla^2 \omega -\frac{3}{2}\sigma\epsilon\nabla\times\mathcal (\nabla^2\phi\nabla\phi) \nonumber \\
    &-& \tau\nabla^2\phi  -\beta\omega\,;\label{eq:omega} \\
    \nabla \cdot \bm u &=& 0\,; \quad \bm \omega = (\nabla \times \bm u) = - \nabla^2 \psi\,;\label{eq:incom} 
    \end{eqnarray}
 the scalar order parameter $\phi$, which distinguishes between regions with rotors that rotate counterclockwise ($\phi > 0$) and clockwise ($\phi < 0$), is coupled to the fluid velocity $\bm u$ as in the incompressible Cahn-Hilliard-Navier-Stokes PDEs [$\tau=0$ in Eq.~\eqref{eq:omega}] for a binary-fluid mixture~\cite{pal2016binary,perlekar2017two,padhan2024novel} with the free-energy functional $\mathcal F$, 
 the constant fluid density $\rho = 1$, and $M$, $\epsilon$, $\sigma$, $\nu$, and $\beta$ the mobility, interfacial width, surface tension, kinematic viscosity, and friction coefficient, respectively. The strength of the activity depends on the magnitude of the torque $\bm{\tau} = \tau \hat{e}_z$, which is perpendicular to the $xy$ plane like the vorticity $\bm{\omega} = \omega \hat{e}_z$ that is related to the stream function $\psi$ via the Poisson equation~\eqref{eq:incom}.

 For the initial condition, we use a statistically homogeneous state with $\phi(x,y,t=0)$ independent and identically distributed random numbers drawn uniformly from the interval $[-0.1, 0.1]$; we set the vorticity $\omega(x,y,t=0)$ to zero. 
The statistical properties of the nonequilibrium, statistically steady state (NESS) of active-rotor turbulence depend on the Reynolds number $Re \equiv L u_{rms}/{\nu}$, the Cahn number $Cn \equiv \epsilon/L $, the Weber number $We \equiv L u_{rms}^2/\sigma$, the P\'eclet number $ Pe \equiv Lu_{rms} \epsilon/M\sigma$, and the non-dimensionalised activity $\alpha=\tau/u_{rms}^2$ and friction $\beta'=\beta L /u_{rms}$, where $u_{rms}$ is the root-mean-square velocity, and $L$ and $L_{\phi}$ are, respectively, the integral and coarsening-arrest lengths scales, which are defined in terms of the energy spectrum $E(k)$ and phase-field spectrum $S(k)$ [see the Appendix] as follows: 
\begin{eqnarray}
   L&= & \frac{\sum_k k^{-1}E(k)}{\sum_k E(k)} ; \quad L_{\phi} = \frac{\sum_k S(k)}{\sum_k k S(k)}\,; \label{eq:lengths} 
\end{eqnarray}
here, $k$ denotes the wave number.

We use the integral-scale frequency $\Omega_L \equiv u_{rms}/L$ to non-dimensionalize time. 
We monitor the flow topology via the Okubo-Weiss parameter~\cite{okubo1970horizontal,weiss1991dynamics,perlekar2009statistically,pandit2017overview} 
\begin{equation}
\Lambda(x,y) \equiv \frac{\bm{\omega}^2(x,y) - \bm{\Sigma}^2(x,y)}{8}\,,
\label{eq:Okubo}
\end{equation}
where $\bm{\Sigma}$ is the symmetric part of the velocity derivative tensor; $\Lambda > 0$ ($\Lambda < 0$) in the vortical (extensional) regions of the flow.
The spectral energy balance is given by ~\cite{padhan2024novel, boffetta2012two,verma2019energy}
\begin{eqnarray}
\partial_t{{E(k,t)}} &=& - T^u(k,t) - S^{\phi}(k,t) - 2\nu k^2 E(k,t) \nonumber \\  &+& T^{rot}(k,t)  - 2 \beta E(k,t)\,, \label{eq:spectralbalance}
\end{eqnarray}
 where the $T^u(k,t)$, $T^{rot}(k,t)$, $S^{\phi}(k,t)$, $2\nu k^2 E(k,t)$, and $2\beta E(k,t)$ are the $k$-shell averaged contributions from the advective, torque, stress-tensor ($\nabla^2\phi\nabla\phi$), viscous dissipation, and friction, terms, respectively; their time averages in the NESS, $E(k)$, $S(k)$, $T^u(k)$, $T^{rot}(k)$, and $S^{\phi}(k)$,  are defined in Eqs.~\eqref{eq:Ek}-\eqref{eq:Tketc} in the Appendix. 
 
 We solve the PDEs~\eqref{eq:phi}-\eqref{eq:incom} by using pseudospectral direct numerical simulations (DNSs)~\cite{canuto2007spectral,padhan2024novel}, which we describe in the Appendix.  

\begin{figure*}[htp]
\centering
    \includegraphics[width=1.0\textwidth]{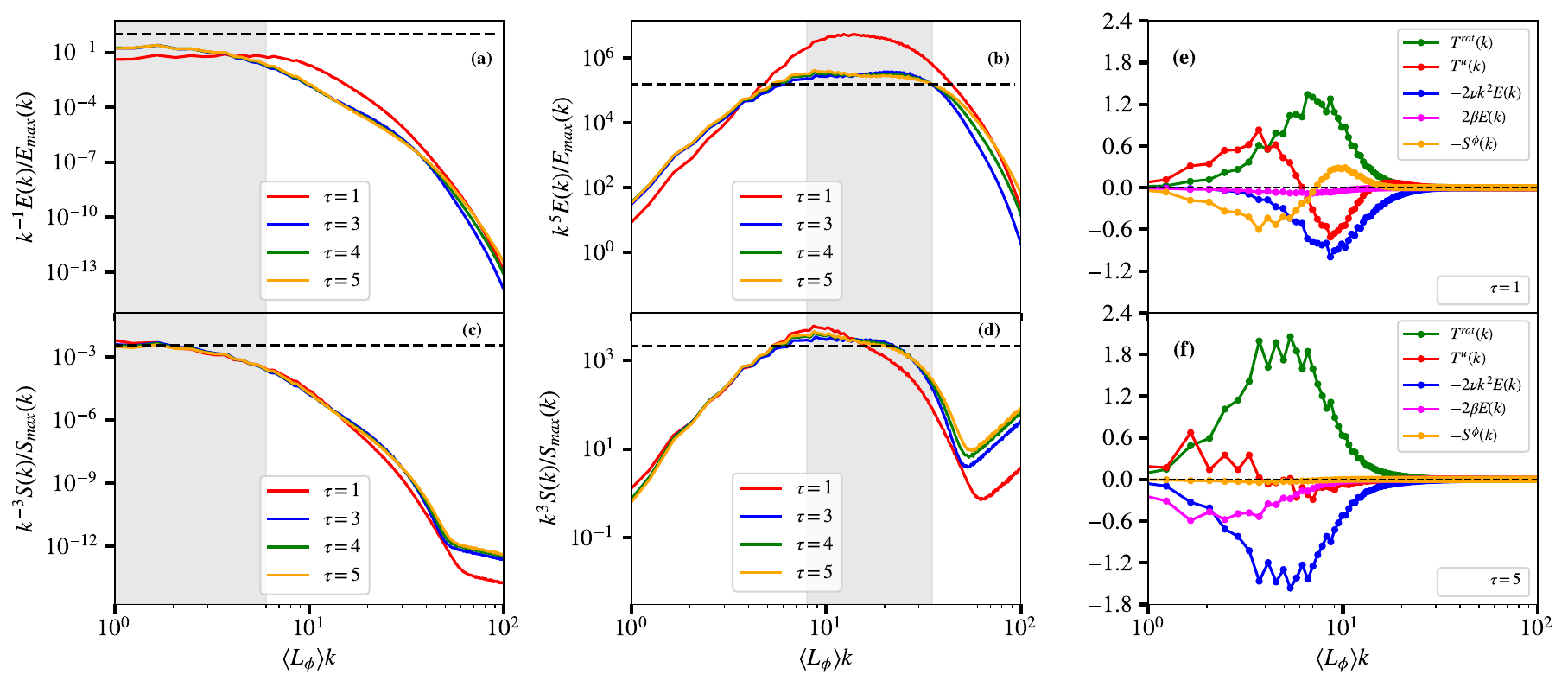}
    \caption{Log-log plots of compensated spectra versus the scaled wavenumber $\langle L_{\phi}\rangle k$, for $\tau=1,\,3,\,4$, and $5$, with power-law scaling regions shaded in gray: (a) $k^{-1}E(k)$ [scaling region $1 \lesssim k \lesssim 8$]; (b) $k^{5}E(k)$ [scaling region $8 \lesssim k \lesssim 35$]. (c) $k^{-3}S(k)$ [scaling region $1 \lesssim k \lesssim 8$]; (d) $k^{3}S(k)$ [scaling region $8 \lesssim k \lesssim 35$]; for the spatiotemporal evolution of these pseudocolor plots, see the Videos V1, V2, V3, and V4 in the Appendix. Log-lin plots of $T^u(k)$(red), $T^{rot}(k)$(green), $ S^{\phi}(k)$(orange), $2\nu k^2 E(k)$(blue) and $2\beta E(k)$( magenta) [see Eq.~\eqref{eq:spectralbalance}] versus $\langle L_{\phi}\rangle k$, in the NESS, for $\beta=0.3$ and (e) $\tau=1$ and (f) $\tau=5$. For plots of the energy and rotational fluxes, $\Pi^u(k)$ and  $\Pi^{rot}(k)$, see Figs.~\ref{fig:pdf_app} (a)-(c) in the Appendix.}
     \label{fig:energy_spectra}
\end{figure*}
\begin{figure*}
    \centering
    \includegraphics[width=1.0\linewidth]{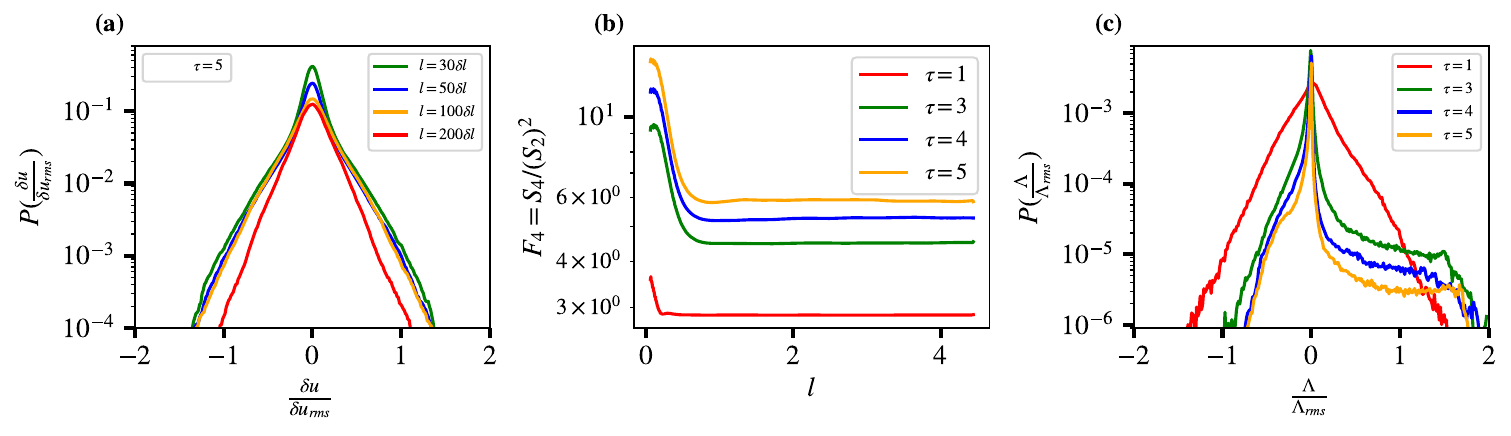}
    \caption{Semilog plots: (a) PDFs of longitudinal-velocity increments $\delta {\bm u}(l)$ for separations $l = 30\delta l$, $l = 50\delta l$, $l = 100\delta l$, and $l = 200\delta l$ for $\tau=5$ [for $\tau=1$, see Fig.~\ref{fig:pdf_app} (f) in the Appendix]; (b) the flatness $F_4 \equiv S_4/(S_2)^2$  versus $l$, for $ \tau = 1, 3, 4,$ and $ 5$,
    shows distinct deviations from the Gaussian value of $3$ for $\tau >1$ and small length scales $l$; (c) the PDF of the Okubo–Weiss parameter $\Lambda$ [see text] for $\tau = 1, 3, 4,$ and $ 5$.}
    \label{fig:pdf_levelc}
\end{figure*}
\section{Results}
We carry out a series of DNSs that we have designed to illustrate the emergence of active-rotor turbulence, as we increase $\tau$ in the ARCHNS model~\eqref{eq:phi}-\eqref{eq:incom}. In Fig.~\ref{fig:pcolor} (b) we present a pseudocolor plot of $\omega$, at a representative time in the turbulent NESS, that contains several vortex doublets [one is enlarged in Fig.~\ref{fig:pcolor} (c)]. Figures~\ref{fig:pcolor} (d), (e), (f), and (g) show how such pseudocolor plots of $\omega$ evolve as we increase $\tau$ from $0.5,\,1.5,\,4$, up until $5$; Figs.~\ref{fig:pcolor} (h), (i), (j), and (k) are, respectively, the counterparts of these plots for $\phi$. [The spatiotemporal evolution of these plots is given in Videos V1, V2, V3, and V4 in the Appendix.]

From Figs.~\ref{fig:pcolor} (e) and (i) we observe that phase separation occurs at $\tau = 0.5$, but it is arrested to some extent. The plots versus $t\Omega_L$ of $E(t)/E(t=1/\Omega_L)$ and $L_\phi(t)$ in Figs.~\ref{fig:length_scale} (a) and (b), respectively, show that the fluctuations in the energy increase with $\tau$, whereas the coarsening-arrest length decreases. We quantify coarsening arrest by the plots in Fig.~\ref{fig:length_scale} (c) which show that the mean domain size $\langle L_\phi \rangle$ (red curve) decreases with $\tau$ as the integral-scale Reynolds number $Re$ (blue curve) increases. 

The pseudocolor plots in the second and third rows of Fig.~\ref{fig:pcolor} uncover a remarkable transition that occurs at $\tau \simeq 1.5$ from a NESS with partial phase separation to another NESS in which active-rotor-induced turbulence (i) suppresses phase separation and (ii) the field $\omega \propto \phi$. This suppression of phase separation, or coarsening arrest, is similar to its counterpart in binary-fluid turbulence~\cite{perlekar2014spinodal,perlekar2017two}. The intriguing proportionality of $\omega$ and $\phi$ has not been seen in any other binary-fluid model. We develop a theory of this proportionality below.

In Fig.~\ref{fig:energy_spectra} we present log-log plots of compensated energy and phase-field spectra versus the scaled wavenumber $\langle L_{\phi}\rangle k$ for $\tau=1,\,3,\,4$, and $5$. The plots in Figs.~\ref{fig:energy_spectra} (a), (b), (c), and (d), which display $k^{-1}E(k)$, $k^{5}E(k)$, $k^{-3}S(k)$, and $k^{3}S(k)$, respectively, are consistent with the following scaling forms at small and intermediate values of $k$ [see the dashed horizontal lines in  Figs.~\ref{fig:energy_spectra} (a)-(d)]: (i) $E(k) \sim k$ and $S(k) \sim k^3$, for $1 \lesssim \langle L_{\phi}\rangle k \lesssim 8$ [especially for $\tau=1$]; and (ii) $E(k) \sim k^{-5}$ and $S(k) \sim k^{-3}$, for $11 \lesssim \langle L_{\phi}\rangle k \lesssim 35$ [especially for $\tau > 1$]. If we assume $\omega \propto \phi$, then the spectrum of $\phi$ is proportional to the enstrophy spectrum, i.e.,  $S(k) \sim \Omega(k) \sim |\tilde{\bm \omega}({\bf{k}})|^2 \sim k^2E(k)$, which is consistent with Fig.~\ref{fig:energy_spectra}. 

In Figs.~\ref{fig:energy_spectra} (e) and (f) we present, for $\tau=1$ and $\tau=5$, respectively, log-lin plots of $T^{u}(k)$, $T^{rot}(k)$, $ S^{\phi}(k)$, $2\nu k^2 E(k)$ and $2\beta E(k)$ [the  Eq.~\eqref{eq:spectralbalance}] versus $\langle L_{\phi}\rangle k$ for the illustrative value $\beta=0.3$. These plots show that, at large times and especially at large values of $\tau$, the dominant terms in the spectral energy balance [see Eq.~\eqref{eq:spectralbalance}] are the active-stress term $T^{rot}(k)$, the dissipation term $2\nu k^2 E(k)$, and the friction term $2\beta E(k)$, which arise, respectively, because of the $\tau\nabla^2\phi$, $\nu\nabla^2\omega$, and $\beta \omega$ terms in Eq.~\eqref{eq:omega};  the torque-driven motion dominates over forces that would normally lead to coarsening. Finally, a dominant-balance argument [see the Appendix] yields the theoretical prediction 
\begin{eqnarray}
   (\nu -  L^2\beta )\omega &\sim& \tau \phi\,;  
   \label{eq:phipropomega}
\end{eqnarray}
this is consistent with our qualitative suggestion $\omega \propto \phi$ in the large-$\tau$ plots of Fig.~\ref{fig:pcolor}. We verify the relation~\eqref{eq:phipropomega} explicitly by making scatter plots of $\omega$ versus $\phi$ for different values of $\tau$ [see Figs.~\ref{fig:pdf_app} (d) and (e) in the Appendix].

Given that the energy spectra in Fig.~\ref{fig:energy_spectra} exhibit power-law scaling ranges that are qualitatively reminiscent of turbulence in 2D fluids~\cite{perlekar2009statistically,boffetta2012two,pandit2017overview} and active fluids~\cite{bratanov2015new,alert2022active,mukherjee2023intermittency,kiran2023irreversibility}, it is natural to ask whether active-rotor 
turbulence also exhibits intermittency and flow topologies of the types seen in classical-fluid~\cite{perlekar2009statistically,boffetta2012two,pandit2017overview} and active-fluid~\cite{alert2022active,mukherjee2023intermittency,kiran2023irreversibility,kiran2025onset} turbulence. To answer this question, we calculate several probability distribution functions (PDFs). The PDFs of the Cartesian components of $\bm u$ are nearly Gaussian  
like their fluid and active-turbulence counterparts [see, e.g., Refs.~\cite{batchelor1953theory,pandit2009statistical,dunkel2013fluid}]. To uncover intermittency, we first define the longitudinal velocity increments $\delta {\bm u}(l)\equiv \big({\bm u}({\bm x}+{\bm l})- {\bm u}({\bm x})\big)\cdot \hat {\bm  l}$ and then obtain their PDFs, for different values of the separation $l= |{\bm l}|$, 
which we display in Fig.~\ref{fig:pdf_levelc} (a) for $\tau =5$ [for $\tau =1$ , see Fig.~\ref{fig:pdf_app} (f) in the Appendix]; these PDFs show distinct scale dependence, more so for $\tau =5$ than for $\tau = 1$, because there is more rotor-induced turbulence in the former case. To quantify this scale dependence, we compute the order-$p$ longitudinal-velocity structure functions $S_p(l) \equiv \langle [\delta {\bm u}(l)]^p \rangle$ and, therefrom, the flatness $F_4(l) \equiv S_4(l)/[S_2(l)]^2$, which increases as $l$ decreases [see Fig.~\ref{fig:pdf_levelc} (b)], a clear signature of small-scale intermittency; the value of $F_4(l)$ increases as $\tau$ goes up from $1$ to $5$; and at large $l$ it is close to $3$, the value of the flatness for a Gaussian PDF, especially for $\tau = 1$.

To investigate the topology of active-rotor flow, we calculate  $\Lambda$, the Okubo-Weiss~\cite{okubo1970horizontal,weiss1991dynamics,perlekar2009statistically,pandit2017overview} parameter~\eqref{eq:Okubo}, and 
plot the PDFs $\mathcal{P}(\Lambda)$ in Fig.~\ref{fig:pdf_levelc} (c) for $\tau = 1,\,3,\,4,$ and $5$; it has zero mean, by definition, and, in the NESS of rotor-induced turbulence [$\tau > 1$], it shows a characteristic cusp at $\Lambda = 0$. We note that $\mathcal{P}(\Lambda)$ is skewed like its counterparts in 2D fluid turbulence~\cite{perlekar2009statistically,pandit2017overview} and in the 2D Toner-Tu-Swift-Hohenberg model for bacterial turbulence~\cite{kiran2023irreversibility,mukherjee2023intermittency}. The skewness of $\mathcal{P}(\Lambda)$ increases markedly as $\tau$ goes from $1$ to $5$.

\section{Conclusions}
We have carried out the first study of active-rotor-induced turbulence in a system of counter-rotating spinners. Using the ARCHNS model~\eqref{eq:phi}-\eqref{eq:incom}, we have demonstrated that, as we increase the activity $\tau$, phase separation into clockwise- and anticlockwise-rotating states is suppressed as active-rotor turbulence emerges. 
This turbulence has several intriguing properties: It displays vortex doublets, which also appear in the absence of turbulence~\cite{sabrina2015coarsening}; next, $\omega \propto \phi$, a remarkable proportionality that has not been seen in any binary-fluid system so far; energy and concentration spectra show well-developed power-law ranges, with $\tau$-independent exponents for $1.5 \lesssim \tau $.  We hope that our work will lead to experimental studies of active-rotor turbulence in both biological~\cite{riedel2005self,petroff2015fast,drescher2009dancing} and synthetic systems~\cite{kokot2018manipulation,soni2019odd,han2020emergence}. 

What could be the biological implications or benefits of such active-rotor turbulence?  Specific answers to these questions can only emerge from new studies. We note, however, that Ref.~\cite{drescher2009dancing} has already conjectured that, ``\ldots flows driving \textit{Volvox} clustering at surfaces enhance the probability of fertilization during the sexual phase of their life cycle.'' 

Our work lays the foundation for new theoretical studies of exotic states in spinner systems that are described by the ARCHNS model~\eqref{eq:phi}-\eqref{eq:incom}. In particular, as we will show elsewhere, this model  can also exhibit states with vortex triplets; such triplets have been obtained in the single-phase study of Ref.~\cite{van2023spontaneous}. 

\begin{acknowledgments}
		We thank V.K. Babu, K.V. Kiran, E. Knobloch, and A. Jayakumar  for discussions, the Anusandhan National Research Foundation (ANRF), the Science and Engineering Research Board (SERB), and the National Supercomputing Mission (NSM), India, for support,  and the Supercomputer Education and Research Centre (IISc), for computational resources. The authors would like to thank the Isaac Newton Institute for Mathematical Sciences, Cambridge, for support and hospitality during the programme \textbf{Anti-diffusive dynamics: from sub-cellular to astrophysical scales}, where some of the work on this paper was undertaken; this work was supported by EPSRC grant EP/R014604/1.
	\end{acknowledgments}
\section{Appendix}
\label{sec:Appendix}
\setcounter{equation}{0}
\renewcommand{\theequation}{A\arabic{equation}}

\subsection{Model} The active-rotor Cahn Hilliard Navier Stokes (ARCHNS) model is governed by the partial differential equations~\eqref{eq:phi}-\eqref{eq:incom}, which use the vorticity form~\eqref{eq:omega} for the Navier-Stokes equation (NSE) in 2D. The velocity form of the NSE is (in both 2D and 3D)
\begin{eqnarray}
     \partial_t \bm u + (\bm u \cdot \nabla) \bm u &=& -\nabla p + \nu \nabla^2 \bm u -\frac{3}{2}\sigma\epsilon (\nabla^2\phi\nabla\phi) \nonumber \\    
     &-& \nabla\times\mathcal(\bm \tau\phi)  -\beta\bm u\,,\label{eq:velocity} 
    \end{eqnarray}
    where $p$ is the pressure; this can be removed by using the incompressibility condition~\eqref{eq:incom} and, if required, the pressure can then be calculated by solving a Poisson equation.
    The strength of the activity depends on the magnitude of the torque $\bm{\tau}$. 
    
    In 2D $\bm{\tau} = \tau \hat{e}_z$, which is perpendicular to the $xy$ plane. The nondimensional form the ARCHNS equations~\eqref{eq:phi}-\eqref{eq:incom} is:
    \begin{eqnarray}
         \partial_t \phi + (\bm u \cdot \nabla) \phi &=& \frac{3}{2Pe}\nabla^2 \left( \frac{1}{2}(\phi^3 -\phi) -Cn^2 \nabla^2 \phi \right)\,; \label{eq:phi-non} \nonumber \\
          \partial_t \omega + (\bm u \cdot \nabla) \omega &=& \frac{1}{Re} \nabla^2 \omega -\frac{3}{2}\frac{Cn}{We}\nabla\times\mathcal (\nabla^2\phi\nabla\phi) \nonumber \\ 
    &-& \alpha\nabla^2\phi  -\beta'\omega\,.\label{eq:omega-non} 
    \end{eqnarray}
\subsection{Numerical Methods}
We solve the PDEs~\eqref{eq:phi}-\eqref{eq:incom} using a square domain of size $2\pi \times 2\pi$, with periodic boundary conditions in both $x$ and $y$ directions, and a pseudospectral direct numerical simulation (DNS)~\cite{canuto2007spectral,padhan2024novel} with $N^2$ collocation points, which evaluates derivatives in Fourier space and products in physical space, and uses the $N/2$ rule for dealiasing. For time integration, we employ the semi-implicit ETDRK-2 method~\cite{cox2002exponential}. Our CUDA C code is optimized for recent GPU processors.
The parameter values that we use in our DNS are as follows: $N=1024$, $M=0.0001$, $\epsilon=0.01839$, $\sigma=1$, $\nu=0.01$, and $\beta=0.3$, whence we obtain the nondimensional parameters listed in Table~\ref{tab:param}.
\begin{figure*}[htp]
   \includegraphics[width=1.0\linewidth]{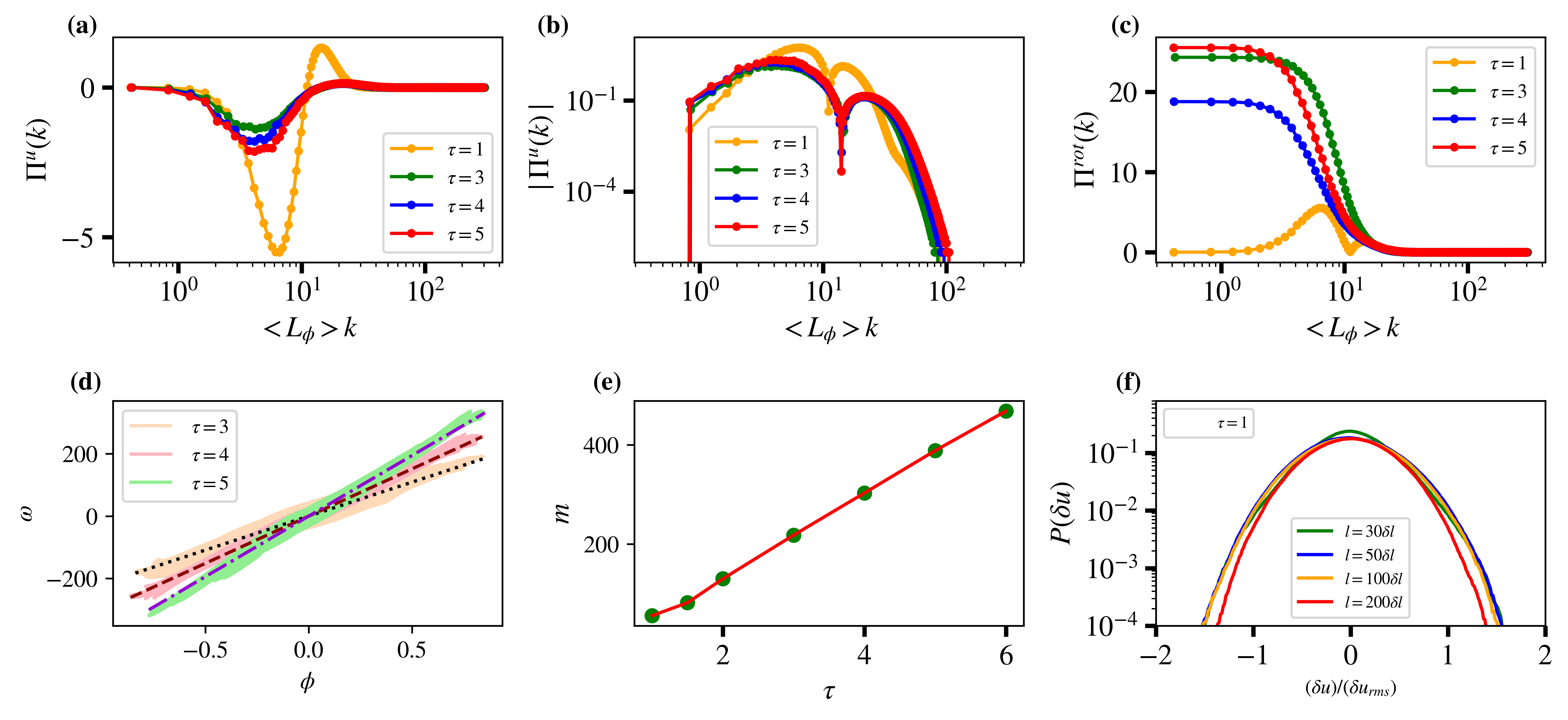}
    \caption{(a) Semi-log plot versus $k$ of the energy flux $\Pi^u(k)$ for the advection term ($\bm u\cdot\grad \bm u$) for $\tau=1,\,3,\,4$, and $5$; (b) log-log plots of $|\Pi^{u}(k)|$  for the values of  $\tau$ in (a). (c) Semi-log plot versus $k$ of the energy flux $\Pi^{rot}(k)$ for the active-stress term ($\tau \grad^2 \phi$) with $\tau=1,\,3,\,4$, and $5$.  (d) Scatter plots of $\omega$ versus $\phi$ illustrating $\omega \propto \phi$, for $\tau=3$ (peachpuff), $\tau=4$ (ligh pink), and $\tau=5$ (light green); the dashed lines show linear fits with slopes $m(\tau)$, which are plotted versus $\tau$ in (e). Semi-log plots (f) of the PDFs of the longitudinal-velocity increments $\delta{\bm{u}}(l)$ for $\tau=1$ and separations $l = 30\delta l$, $l = 50\delta l$, $l = 100\delta l$, and $l = 200\delta l$; these PDFs are nearly Gaussian, unlike those for $\tau=5$ [cf. Fig.~\ref{fig:pdf_levelc} (a)].}
    \label{fig:pdf_app}
\end{figure*}
\subsection{Spectra and transfer terms} The energy and phase-field spectra are defined, respectively, as follows:
\begin{eqnarray}
   \hspace{-0.5cm}E(k,t)&=&\frac{1}{2}\sum_{k'= k-1/2}^{k'= k+ 1/2}|\tilde{\bm{u}}(\textbf{k}',t)|^2; \; E(k) \equiv \langle E(k,t) \rangle_t\,;\label{eq:Ek} \nonumber \\
   \hspace{-0.5cm}S(k,t)&=&\frac{1}{2}\sum_{k'= k-1/2}^{k'= k+ 1/2}|\tilde{\phi}(\textbf{k}',t)|^2; \; S(k) \equiv \langle S(k,t) \rangle_t\,; \label{eq:Sk}
   \end{eqnarray}
   the tildes denote spatial Fourier transforms, $k$ and $k'$ are the moduli of the wave vectors $\mathbf{k}$ and $\mathbf{k}'$, and $<\cdot>_t$ denotes the time average over the statistically steady state.
 The terms $T^u(k)$, $T^{rot}(k)$, and $S^{\phi}(k)$ [in Eq.~\eqref{eq:spectralbalance} in the main paper] are defined as follows: 
\begin{eqnarray}
    T^u(k)&=&\sum_{k'= k-1/2}^{k'= k+ 1/2} \langle\widetilde{\textbf{u}(-\textbf{k}'}).\textbf{P}(\textbf{k}').\widetilde{(\textbf{u}.    \grad\textbf{u})}(\textbf{k}')\rangle_t \,;\nonumber  \\
    T^{rot}(k)&=&\sum_{k'= k -1/2}^{k'= k +1/2}  \langle\widetilde{\textbf{u}(-\textbf{k}'}).(\widetilde{\nabla \cp \vb{\tau}\phi})(\textbf{k}')\rangle_t\,;\nonumber  \\
    S^{\phi}(k)&=&\sum_{k'= k-1/2}^{k'= k+ 1/2} \langle\widetilde{\textbf{u}(-\textbf{k}'}).\textbf{P}(\textbf{k}').\widetilde{(\nabla^2\phi\nabla\phi)}(\textbf{k}')\rangle_t \,;\nonumber\\
    \Pi^u(k)&=&\sum_{k'=0}^{k'=k} T^u(k')\,; \nonumber\\
    \Pi^{rot}(k)&=&\sum_{k'=0}^{k'=k} T^{rot}(k');\; \label{eq:Tketc}
\end{eqnarray}
\begin{table}[h]
    \centering
    \renewcommand{\arraystretch}{1.2} 
\begin{tabular}{ |c|c|c|c|c|c|c|} 
 \hline
  $\tau$ & $\alpha$ & $\beta'$  & $Cn$ & $We$ & $Pe$ & $Re$  \\ 
 \hline
 0.5 & 0.13933   & 0.01664 & 0.18489 &  0.35725 & 34.71577 & 18.85930 \\ 
  \hline
 1 & 0.19799   & 0.00914 & 0.23925 &  0.38857 & 31.8279569 & 17.29050 \\ 
 \hline
 1.5 & 0.28199  & 0.01313 & 0.15811 & 0.61925 & 49.42520  & 26.85018 \\ 
 \hline
 3 & 0.33407  &  0.00853 & 0.10802 & 2.04034 & 108.54162  & 58.96511 \\ 
 \hline
 4 &  0.24609 & 0.00617 & 0.11004 & 2.71882 & 124.13768 &  67.43765 \\
 \hline
 5 & 0.24024 &  0.00467 & 0.11341 & 3.37768 & 136.29100 &  74.03993 \\ 
 \hline
 \end{tabular}
 \caption{Table of the values of the non-dimensional parameters, for different values of $\tau$ [column 1]: Cahn number $Cn \equiv \epsilon/L $, Weber number $We \equiv L u_{rms}^2/\sigma$, P\'eclet number $ Pe \equiv Lu_{rms} \epsilon/M\sigma$, and the non-dimensionalised activity $\alpha=\tau/u_{rms}^2$ and friction $\beta'=\beta L /u_{rms}$, where $u_{rms}$ is the root-mean-square velocity.}
 \label{tab:param}

\end{table}
here $P_{ij}(k)=\delta_{ij} -\frac{k_i.k_j}{k^2}$ are the elements of the transverse projector $\textbf{P}(\textbf{k})$; and the fluxes because of advection and active stress are, respectively, $\Pi^u(k)$ and $\Pi^{rot}(k)$.
\subsection{Relation between $\omega$ and $\phi$ at large $\tau$}
The calculations of the terms in the energy budget~\eqref{eq:spectralbalance} can be used to show that $\omega \propto \phi$ at large $\tau$: The active-stress term ($\tau\nabla^2\phi$) and the dissipation term ($\nu\nabla^2\omega$) dominate the energy budget [see Fig.~\ref{fig:energy_spectra} (f) for $\tau=5$].
From Eqs.~\eqref{eq:phi}-\eqref{eq:incom}, dimensional analysis in the statistical steady state reveals that the second and fourth terms, representing the advection term and with the stress-tensor term, contribute negligibly compared to the active-stress, dissipation, and friction terms. Consequently, we can use the following dominant balances:
\begin{eqnarray}
\frac{\phi u}{L} &\sim& M \frac{3}{4}\frac{\sigma}{\epsilon}\frac{1}{L^2} (\phi - \phi^3)\,;\nonumber \\ 
\frac{\omega u}{L} &\sim& \frac{\nu \omega}{L^2}  -\frac{3}{2}\sigma\epsilon\frac{\phi^2}{L^2} - \frac{\tau \phi}{L^2} -\beta \omega \,;\nonumber \label{eq:balance_term}\\
&\Rightarrow& (\nu -  L^2\beta )\omega \sim  \tau \phi\,;
\end{eqnarray}
this is relation~\eqref{eq:phipropomega} in the main paper; and it is verified by the scatter plots of $\omega$ versus $\phi$ in Fig.~\ref{fig:pdf_app} (d).
\subsection{Videos}
The following videos show the spatiotemporal evolution of the pseudocolor plots in Fig.~\ref{fig:pcolor}:
\begin{itemize}
\item Video V1: This shows the spatiotemporal evolution of pseudocolor renderings [cf. Figs. 1 (d) and (h)] of the fields $\phi$ (left panel) and  $\omega$ (right panel) for $\tau=0.5$ (\textcolor{blue}{\url{https://youtu.be/kyjbdYTGsRY}}); 
\item Video V2: this shows the spatiotemporal evolution of pseudocolor renderings [cf. Figs. 1 (e) and (i)] of the fields $\phi$ (left panel) and  $\omega$ (right panel) for $\tau=1.5$ (\textcolor{blue}{\url{https://youtu.be/-2XRSvvGAEE}});
\item Video V3: this shows the spatiotemporal evolution of pseudocolor renderings [cf. Figs. 1 (f) and (j)] of the fields $\phi$ (left panel) and  $\omega$ (right panel) for $\tau=4$ (\textcolor{blue}{\url{https://youtu.be/yKZ3cUNktpg}});
\item Video V4: this shows the spatiotemporal evolution of pseudocolor renderings [cf. Figs. 1 (g) and (k)] of the fields $\phi$ (left panel) and  $\omega$ (right panel) for $\tau=5$  (\textcolor{blue}{\url{https://youtu.be/1-H61YNS88M}}).
\end{itemize}
\bibliography{main}

\end{document}